**Performance Evaluation of Damping Systems in Civil Engineering Structures Via Minimal Sensor**

Xinhao He[a*], Dan Li[b*]

[a] *Graduate School of Engineering, Department of Civil and Environmental Engineering, Tohoku University, Sendai, Japan*
[b] *School of Civil Engineering, Southeast University, Nanjing, China*

*Corresponding author.
Email addresses:* xinhao.he.a8@tohoku.ac.jp (X. He), danli@seu.edu.cn (D. Li).

**Abstract**

To control structural responses under various actions, the growing use of supplementary damping systems in modern civil engineering structures necessitates inspecting and evaluating their operational performance post-installation. However, due to the dispersed placement and complex nonlinearities of these devices, difficulties arise in determining minimal sensor configuration. This is inherently connected to a pivotal challenge: establishing a reliable input-output mapping, which comprises both the mathematical model and sensor arrangements. Prior work indicates this can be achieved through theoretical observability analysis or Lie symmetries analysis, both of which provide different perspectives on the existence of a way to access the solutions of a system identification problem uniquely (at least locally). The present study introduces a unified framework, enhanced by algorithm realization as an application guide, for analyzing the observability and Lie symmetries of a given input-output mapping. We demonstrate its implementation via examples of a building structure with various damping systems under different conditions such as seismic loads, wind loads, and operational vibrations. Finally, we present a case study for an isolation building with an inerter damper and minimal sensor arrangement under seismic action. The results demonstrate that the unscented Kalman filter, a system identification method, can precisely estimate structural responses and assess damping device performance once a reliable input-output mapping is established.

**Keywords**: Damping Systems, Damper, Seismic Isolation, Observability Analysis, Lie Symmetries, Unscented Kalman Filter

Secondary Keywords: Inerter Damper, Nonlinear Energy Sink, Base Isolation, Lead Rubber Bearing, Friction Pendulum Bearing, Viscous Damper

1. **Introduction**

Energy dissipation devices, commonly known as damping systems [1], have found extensive applications in modern civil engineering structures, assisting them in withstanding various external actions such as earthquakes, winds, and traffic-induced vibrations. In general, one or more types of these devices are strategically placed throughout the main structure to maximize energy absorption, thus ensuring the main structure remains at its elastic range. Due to their crucial role, the attention given to their maintenance and performance assessment post-installation has steadily grown. While several approaches exist for evaluating the performance of specific devices under operational conditions, the advent of innovative damping systems necessitates a unified framework for performance assessment.

For instance, viscous dampers [2,3] were first employed in building structures in the 1980s and have since found wide-ranging use in vibration control of diverse civil engineering structures. The concept of the tuned mass damper [4–6], originated from vibration absorbers, divert the energy flow of the main structure to an auxiliary mass that dissipate the energy by vibrating out of the phase with the structure. Steel dampers, such as steel shear panel [7,8] and buckling-restrained brace [9,10], have seen considerable use, providing stable energy dissipation and controllable yield displacement through plastic deformation and device-specific structural design.

Nonetheless, in response to evolving needs, a variety of innovative dampers have been extensively explored in recent years [11,12]. For example, laminated rubber dampers [13] were proposed to achieve large displacement stroke and high energy dissipation capability while retaining cost-effectiveness. Inerter dampers [14] were proposed to provide sufficient damper force by leveraging the rotational motion of a relatively small mass. Rate-independent dampers [15–17], approximating the behavior of complex stiffness model, were designed to prevent large displacement of low-frequency structures subjected to long-period earthquakes. Similar purposes are achieved by negative stiffness dampers [18–20]. Examples of their use can be found at isolation layers of bridges and buildings, or across multiple spans and layers. Some typical dampers and their application in building structures are presented in Fig. 1a.

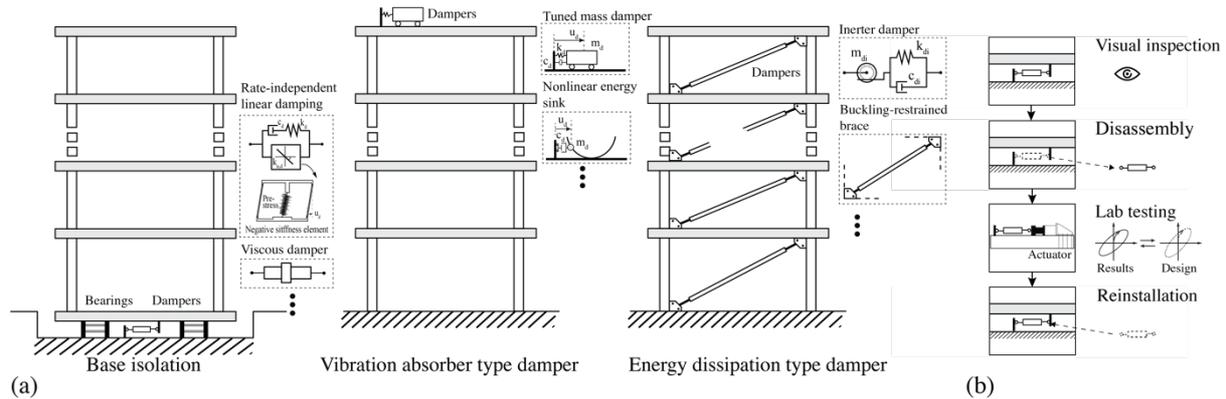

**Figure 1.** (a) Structures with Damping Systems; (b) Current Methodology for Performance Evaluation

From a maintenance standpoint, periodic inspection is necessary for evaluating the performance of these damping systems following installation. This typically entails comparing the dynamic characteristics of the main structure (e.g., natural frequency, modal shape, damping ratio, and measured responses) with the predicted values of an analytical model [21–25]. Although various studies introduce methods for appraising the performance of specific devices, e.g., hybrid base isolation systems [26], friction pendulum bearings [27–29], and negative stiffness damper [30], these evaluations often necessitate visual inspections, disassembly, laboratory testing, and subsequent reinstallation, see Fig. 1b. It is observed that a well-established methodology for evaluating the functionality or serviceability of an individual device or a category of devices under operational conditions remains elusive. This is due in part to the complex nonlinear behavior of some damping devices and the dispersed placement of these devices throughout structures, arising difficulties in reconstructing the contribution of damping devices and determining a reasonable sensor configuration.

The issue aforementioned inherently pertains to a vital aspect of system identification problem: whether a unique solution exists. Theoretical observability analysis serves as a measure of the capacity to uniquely infer, at least in a local context, the states of a given mathematical model from limited measurements. In instances where the observability condition is unmet, there are infinite sets of erroneous state solutions that can satisfy the input-output mapping, yielding unreliable estimation results as reported in Ref. [31,32]. The exploration of observability has been a crucial subject since Kalman's work on linear dynamic systems [33]. A comprehensive review of nonlinear observability methods is available in Chatzis et al. [34]. For instance, the observability rank condition (ORC), as proposed by Hermann and Krener [35] is an algorithm for examining the local weak observability of analytic nonlinear systems. Recent extensions of ORC [36–38] can address some its limitations in specific cases, such as the system being in an affine input form, the measurements not being an input function, and the inputs being fully measured.

Conversely, theoretical observability has been identified as fundamentally intertwined with the Lie symmetries of an input-output mapping, signifying some invariance following certain transformations. Numerous prior studies have focused on computing the expression of Lie symmetries, providing invaluable insights into a mapping and guiding us towards a unique solution in system identification. For instance, Anguelova et al. [39] devised an algorithm for computing the translation, scaling, affine, and quadratic types of Lie symmetries. To overcome the aforementioned limitations pertaining to the type of Lie symmetries, Sedoglavic [40] introduced a method for calculating the general expression of Lie symmetries using the observability matrix with polynomial coefficients. More recently, Shi and Chatzis [41] provided a framework for determining the general expression of Lie symmetries in cases of unknown inputs.

In summary, reconstructing the contribution of damping devices distributed throughout a structure using measurements from specific locations necessitates the successful establishment of a mathematical model and sensor layout. This objective can be accomplished either by analyzing corresponding theoretical observability or Lie symmetries, both of which provide different perspectives on the existence of a way to access the solutions of a system identification problem uniquely (at least locally). Previous studies have shown that when applying to linear dynamic systems, the theoretical observability analysis corresponds to the rank condition of a linear equation group, while the existence of Lie symmetries corresponds to infinitely many solutions of the equation group. Nevertheless, the connotation of observability and Lie symmetries, particularly their algorithmic realization, remains opaque for researchers and engineers in the field of civil engineering.

The paper is structured as follows: Section 2 formulates the problem of assessing damping systems in structures using system identification techniques and specific structural measurements. Section 3 introduces the concept of solution uniqueness for a given input-output mapping under certain transformations. Section 4 elaborates on the relationship between Lie symmetries computation and the observability rank condition. Section 5 presents a unified framework and its algorithmic realization, illustrating the integration of any ORC-based observability algorithms by redefining the Lie derivatives of the output function. Specifically, the computation of the general expression of Lie symmetries is based on Lie's first theorem and the computation of the null space of the Jacobian matrix in observability analysis algorithms. This section also reviews three observability analysis algorithms applicable to problems with affine input, non-affine input, and unknown inputs. Section 6 offers case studies to demonstrate these algorithms, including a case examining different Lie derivatives definitions' consistency, and another detailing sensor requirement changes under various conditions. Finally, we present a study for an isolation building equipped with an inerter damper and minimal sensors under seismic action, illustrating how the unscented Kalman filter can accurately estimate structural responses and assess damping devices' performance once an observable input-output mapping is established.

**List of symbols**

| | |
|---|---|
| A, B, C | Constant matrices of continuous linear dynamic systems |
| $A_d, C_d$ | Constant matrices of discrete linear dynamic systems |
| $f_t(\cdot), f(\cdot), f^{(n)}(\cdot)$ | System equations for original system, augmented system with unknown parameters, and further augmented system with unmeasured inputs up to order n |
| $f_a(\cdot), g_{uj}(\cdot), h_0(\cdot)$ | Nonlinear vector functions in affine-input form without unmeasured inputs |
| $f_a^{(n)}(\cdot), g_{uj}^{(n)}(\cdot), S_j^{(n)}, h_{uj}(\cdot), h_{wj}(\cdot)$ | Nonlinear vector functions in affine-input form with measured and unmeasured inputs |
| $x_t, x, x^{(n)}$ | State variable vector for original system, augmented system with unknown parameters, and further augmented system with unmeasured inputs up to order n |
| $\theta, t$ | Estimated parameter vector and time |
| $u, w, w^{(n)}$ | Measured inputs and unmeasured inputs, where $w^{(n)} = \frac{d^n w}{dt^n}$ |
| $q, F(\cdot), H(\cdot)$ | General system state vector, system function, and measurement function |
| $y$ | Measurement vector |
| $N_S$ | Dimensions of a vector $S$ |
| $^i\phi$ | ith Lie group of symmetries in an input-output mapping, i=1,2,..,r |
| $^i\xi$ | Infinitesimals of the ith Lie group of symmetries, i=1,2,..,r |
| $\epsilon_i$ | A constant parameter within a continuous interval $S \subseteq \mathbb{R}$ |
| $\Omega_j$ | jth order of the Lie derivatives of the output function |
| $d^{(n)}\Omega$ | Jacobian matrix up to order n |
| $d^{(n)}\Omega_n^m$ | Jacobian matrix by removing the m-th column from $d^{(n)}\Omega_n$ |
| $\mu(\cdot), T(\cdot)$ | Analytic transform functions |
| $z$ | General state vector in the proposed algorithm |
| $\mathcal{O}$ | Observability matrix of linear systems |

**2. Problem Formulation**

The evolution of maintenance in civil engineering requires not only overall structural evaluations but also detailed analysis of individual devices or a category of devices under operational conditions. Yet, installing sensors on every device is impractical and costly. This dilemma leads us to system identification techniques, where the contribution of individual elements is inferred from indirect measurements, given appropriate mathematical models and sensor layouts, as depicted in Fig. 2.

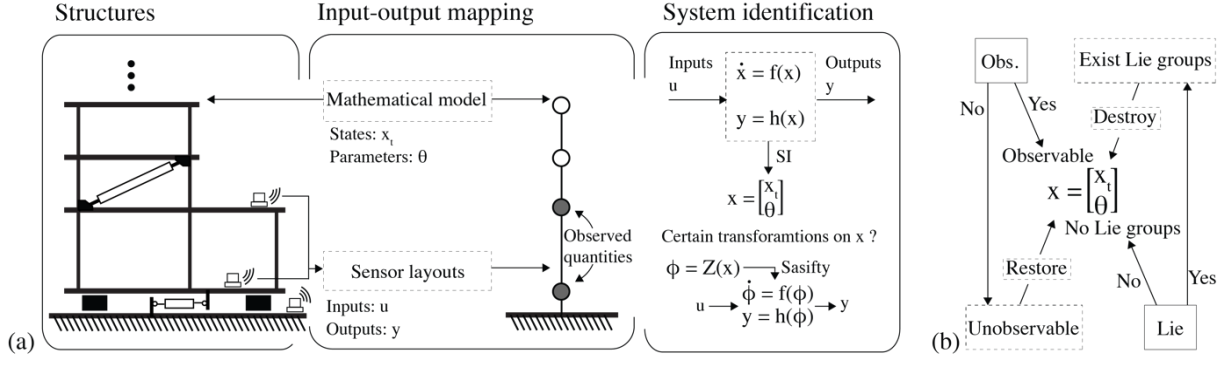

**Figure 2.** (a) Establishment of a Reliable Input-Output Mapping for System Identification; (b) Relationship Between Lie Symmetries and Observability

As shown in Fig. 2a, given sensor layouts and appropriate mathematical model approximation, the input-output mapping of a dynamic system for a system identification problem can be expressed in a state-space form as:

$$\dot{x}_t = f_t(x_t, \theta, u), \qquad y = H(x_t, \theta, u), \qquad \dot{\theta} = 0 \tag{1}$$

where $x_t$ denotes the state vector of the dynamic system, generally made up of displacements and velocities of each degree-of-freedom; $\theta$ denotes parameter vector; $u$ is the input; overdot denotes the derivative with respect to time; and $f_t(\cdot)$ and $H(\cdot)$ are the system and observation functions, respectively. Without the loss of generality, the parameters are assumed to be time-invariant.

If reliable mathematical models can be used to characterize and evaluate the integrity of damping devices, an augmented system can be realized to jointly estimate the unknown parameters as:

$$x = \begin{bmatrix} x_t \\ \theta \end{bmatrix}, \qquad \dot{x} = f(x, u) = \begin{bmatrix} f_t(x_t, \theta, u) \\ 0_{N_\theta \times 1} \end{bmatrix}, \qquad y = H(x, u) \tag{2}$$

where the measurement function remains unvaried, while the system function is augmented to $N_{x_t} + N_\theta$ dimensions. This formulation is versatile since it is tied into two essential objectives in performance assessment of civil engineering structures: structural response prediction and parameter estimation.

Furthermore, there exist instances where partial unmeasured inputs are needed to be estimated, such as wind loads and specific site-effect of earthquakes, the input-output mapping can be further augmented as:

$$x^{(n)} = \begin{bmatrix} x_t \\ \theta \\ w^{(0)} \\ \vdots \\ w^{(n)} \end{bmatrix}, \qquad \dot{x}^{(n)} = \begin{bmatrix} f_t(x, u, w) \\ 0_{N_\theta \times 1} \\ w^{(1)} \\ \vdots \\ w^{(n+1)} \end{bmatrix} = f^{(n)}(x^{(n)}, u, w^{(n+1)}), \qquad y = H(x^{(n)}, u, w^{(n+1)}) \tag{3}$$

where the measurement function still remains unvaried, $w^{(i)}$ denotes the derivative of the unknown input vector with respect to time at i order, and $x^{(i)}$ and $f^{(n)}(\cdot)$ are the corresponding augmented vector and system function, respectively.

Note that all the above formulas are equivalent if the respective state variables remain identical. For brevity, in the following derivations, a general form is used to represent the above three cases.

$$q = x_t \text{ or } x \text{ or } x^{(n)}, \qquad \dot{q} = F(q, u), \qquad y = H(q, u) \tag{4}$$

In addition, the systems discussed in this study are assumed to be smooth and differentiable. For nonsmooth system, Chatzis et al. [34] demonstrated that we can separate it into several analytic branches, yielding to the development of the discontinuous extended Kalman filter [31].

The challenge in establishing a reliable input-output mapping lies in ensuring a unique system identification solution (at least locally). This is achievable via theoretical observability analysis or Lie symmetries computation (see Fig. 2b). The two concepts are complementary and provide insight into whether a unique solution exists (observable), or whether certain transformations lead to invariance in the output, implying non-unique solutions (unobservable). If unobservable, observability needs restoration, requiring the existing Lie symmetries to be destroyed.

However, both concepts remain inaccessible to civil engineers due to the lack of clear algorithm realization. This often demands a deep understanding of numerous complex definitions and intricate mathematical expositions, hindering practical application. Hence, this study seeks to bridge this gap, illuminating the connection between Lie symmetries computation and the observability rank condition, presenting a unified algorithm framework for both computations, and providing practical examples for enhanced understanding.

## 3. The Existence of a Unique Solution in an Input-Output Mapping of a Dynamic System

A successful input-output mapping relies on the existence of a unique solution q for a specified measurement input u and output y. Essentially, there should not exist any transformation on the dependent (q) and independent (time) variables of system f(·) and output h(·) functions that satisfy the input-output mapping (u→y). In this context, Lie symmetry analysis [42–44] offers valuable insight into the invariances of corresponding differential or non-differential equations under variable transformation. Specifically, this transformation forms a local group of point transformation that maps the equation solutions to other solutions.

Herein, we briefly review some essential definitions and properties of Lie symmetries, keeping the symbols consistent with Shi's study [41]. Without losing generality, it is assumed that the input-output mapping of a system contains r (∈ ℕ) groups of Lie symmetries, and each denoted as a one-parameter group of transformations:

$$^i\phi(q,\epsilon_i)^T = \begin{bmatrix} ^i\phi_1 & ^i\phi_2 & \ldots & ^i\phi_{N_\phi} \end{bmatrix}, \qquad i = 1,2,\ldots,r \tag{5}$$

where each element of $^i\phi(q,\epsilon_i)^T$ corresponds to the Lie symmetry of each component of q. The transformation group depends on the single constant parameter $\epsilon_i$ within a continuous interval $S \subseteq \mathbb{R}$. Computing Lie symmetries involves the search of the one-parameter Lie group of transformations leaving the mapping invariant. For instance, $^i\phi_j$ (j = 1,2, … $N_\phi$) could be the component transformed from jth component of q via a one-parameter Lie group of translation ($^i\phi_j = q_j + a_{i,j}\epsilon_i$), scaling ($^i\phi_j = e^{a_{i,j}\epsilon_i}q_j$), inversion ($: ^i\phi_j = \frac{q_j}{1-a_{i,j}\epsilon_i q_j}$), affine, quadradic, polynomial symmetries, and so forth.

In addition to key definitions and theorems provided in Appendix I, we also review here some important properties of Lie symmetries:
1) $^i\phi|_{\epsilon_i=0} = q$;
2) $^i\phi$ is smooth (infinitely differentiable) with respect to the dependent variables q, and is a set of analytic functions of $\epsilon_i$;
3) The following transformations hold for all $a, b \in S$:
$$^i\phi(^i\phi(q,b),a) = ^i\phi(q,\mu(a,b)), \qquad i = 1,2,\ldots,r$$
where $\mu(\cdot)$ is an analytic function of a and b. A one-parameter Lie group of transformations of any other group of the system forms a two-parameter Lie group of symmetries, which still fulfill the mapping.

If r one-parameter symmetries exist, they can be combined to form an r-parameter Lie group of symmetries by applying the Lie group of transformation successively:

$$\phi(q,\epsilon) = ^r\phi\{\ldots, ^2\phi[^1\phi(q,\epsilon_1),\epsilon_2],\ldots,\epsilon_r\} \tag{6}$$

where φ represents the r-parameter Lie group of symmetries with $\epsilon = \{\epsilon_1,\ldots,\epsilon_r\}$. Thus, the presence of Lie symmetries suggests that for any value of ϵ in S and any known input u, φ fulfills the input-output mapping as well as q, leaving the output y unchanged:

$$\frac{d(\phi)}{dt} = F(\phi,u), \qquad y = H(\phi,u) \tag{7}$$

It concludes that the existence of Lie symmetries cannot assure a unique system identification result since multiple sets of solutions for q, corresponding to $^i\phi$ with various value of $\epsilon_i$ as well as $\phi$ with different $\epsilon$ values. Mathematically, the uniqueness of the solutions of a given input-output mapping demands:

$$\phi \equiv q \tag{8}$$

## 4. The Detection of the Uniqueness of Solutions for System Identification

The uniqueness of solutions in a given input-output mapping can be investigated through the detection of the existence of Lie symmetries [39–41]. Shi and Chatzis [41] presented a computational framework for determining the general expression of Lie symmetries in cases of unmeasured input, based on an extended Lie derivative of the output function [36,45]. We demonstrate that this framework can be applied to other cases by redefining the Lie derivative.

The framework aligns with the Lie's First fundamental theorem (see Appendix I), capable of identifying the general expression of Lie symmetries by solving the following first order differential equation with given initial values:

$$\frac{d(^i\phi)}{d\epsilon_i} = {^i\xi}(^i\phi), \qquad {^i\phi}(q,\epsilon_i)|_{\epsilon_i=0} = q \tag{9}$$

where $^i\xi|_{\epsilon_i=0}$ are defined as the infinitesimals of the ith Lie group of symmetries.

It is assumed that the system contain r one-parameter Lie groups of symmetries so that each Lie group fulfills the input-output mapping as:

$$\frac{d(^i\phi)}{dt} = F(^i\phi, u), \qquad y = H(^i\phi, u), \qquad i = 1,2,\dots,r \tag{10}$$

Let us differentiate both sides of the above output function with respect to time at order j:

$$y^{(j)} = \frac{d^j[H(^i\phi, u)]}{dt^j}, \qquad j = 0,1,\dots,n \tag{11}$$

Since the above relationship is independent of $\epsilon_i$, differentiating the equation with respect to $\epsilon_i$ leads to:

$$\frac{dy^{(j)}}{d\epsilon_i} = \frac{d\left\{\frac{d^j[H(^i\phi, u)]}{dt^j}\right\}}{d\epsilon_i} = 0, \qquad j = 0,1,\dots,n \tag{12}$$

The chain rule can be applied to the above expression and combined with Lie's First fundamental theorem, see Eq. (9), as:

$$\frac{\partial\left\{\frac{d^j[H(^i\phi, u)]}{dt^j}\right\}}{\partial(^i\phi)}\frac{\partial(^i\phi)}{\partial\epsilon_i} = \frac{\partial\left\{\frac{d^j[H(^i\phi, u)]}{dt^j}\right\}}{\partial(^i\phi)}{^i\xi}(^i\phi, u) = 0 \tag{13}$$

where $j = 0,1,\dots,n$, and $i = 1,2,\dots,r$.

Based on the first property of Lie symmetries, $^i\phi|_{\epsilon_i=0} = q$, the above equation can be evaluated at $\epsilon_i = 0$ as:

$$\frac{\partial\left\{\frac{d^j[H(q, u)]}{dt^j}\right\}}{\partial(q^{(n)})}{^i\xi} = 0, \qquad j = 0,1,\dots,n \tag{14}$$

Specifically, by applying the chain rule, the jth order of the output function with respect to time can be expressed as:

$$\frac{d^j y}{dt^j} = \frac{d\left(\frac{d^{j-1}y}{dt^{j-1}}\right)}{dt} = \frac{d\left(\frac{d\left(\frac{d^{j-2}y}{dt^{j-2}}\right)}{dt}\right)}{dt} = \frac{d\left\{\cdots \frac{d\left[\frac{d\left(\frac{dy}{dt} = \frac{dy}{dq}\frac{dq}{dt} = \frac{dy}{dq}F = \Omega_1\right)}{dt} = \frac{d(\Omega_1)}{dq}\frac{dq}{dt} = \Omega_2\right]}{dt}\right\}}{dt} = \Omega_j \quad (15)$$

where $\Omega_j$ denotes the Jacobian of the jth order of the Lie derivatives of the output function.

Therefore, Eq. (14) can be expressed as:

$$\frac{\partial(\Omega_j)}{\partial(q)}{}^i\xi = 0, \qquad j = 0,1,\ldots,n, \qquad i = 1,2,\ldots,r \quad (16)$$

Taking all $j = 0,1,\ldots,n$ into consideration, one attains the following homogeneous equation:

$$\begin{bmatrix}\frac{\partial(\Omega_0)}{\partial(q)}\\ \frac{\partial(\Omega_1)}{\partial(q)}\\ \vdots \\ \frac{\partial(\Omega_n)}{\partial(q)}\end{bmatrix}{}^i\xi = [d^{(n)}\Omega]{}^i\xi = 0, \qquad i = 1,2,\ldots,r \quad (17)$$

It is seen that the vector of the infinitesimals can be expressed as column elements in the null space of the Jacobian matrix $d^{(n)}\Omega$:

$${}^i\xi = \text{null}(d^{(n)}\Omega), \qquad i = 1,2,\ldots,r \quad (18)$$

By substituting the expression of the above infinitesimals into Eq. (9) and solving the first order differential equation, the general expression of Lie symmetries, ${}^i\phi$, $i = 1,2,\ldots,r$, can be obtained. In addition, an algorithm using power series to approximate the ith Lie group of the general expression of Lie symmetries can be found in Ref. [41].

Conversely, the existence of a unique solution implies the absence of Lie symmetries and ${}^i\xi$. This links to the rank condition for the Jacobian matrix, see Eq. (19), which is exactly the observability rank condition [34,36,37,45–47].

$$\text{rank}(d^{(n)}\Omega) = \dim(q) \quad (19)$$

## 5. Algorithm realization and application guide

*5.1 Algorithms*

Equations (17) and (19) provide valuable insight into the relationship between the existence of Lie symmetries and the observability rank condition. The focal point lies in the definition of the Lie derivative of the output function, which varies depending on the form of the established input-output mapping.

Table I revisits three definitions of the Lie derivatives of the output function, each capable of addressing different situations in civil engineering. In particular, if the input-output mapping is expressed as an affine-input form without input terms in the output function, the definition was provided in Isidori's study [46] and thoroughly reviewed in Chatzi's study [34], see Table I (1). If both the system and output functions are expressed as an affine-input form with measured or unmeasured inputs, Maes [37] provides a definition in Table I (2), which expanded upon Martinelli's definition [47]. If the input-output mapping is expressed as a general form with measured or unmeasured inputs, Kalsson' [45] and Shi' [36] studies provide the corresponding definition, see Table I (3). Essentially, all these studies focus on the observability analysis of the established input-output mapping.

Table II presents a unified algorithm realization for both the computation of the general expression of Lie symmetries and the observability analysis. One of the key advantages of Lie's theory is that its application is entirely algorithmic, despite some calculations being potentially cumbersome and tedious. Symbolic toolboxes in software like MATLAB, Mathematica, and Maple can be used without having to predetermine specific values for variables. As a result, the algorithm is versatile and capable of dealing with various system constraints such as analytic function restrictions of geometric nonlinearities and rational function restrictions of underlying linear structures.

**Table II. Unified Algorithm Realization**

| Common part | |
| --- | --- |
| Initial | Set $k_{max}$, n = 0, z = x or $x^{(0)}$ (see Table I), $\Omega_0$ (see Table I), and $d\Omega_0 = \frac{\partial \Omega_0}{\partial z}$ |
| Step 1 | n = n + 1 |
| Step 2 | Compute $\Omega_n$ (see Table I) |
| Step 3 | Update z = x or $x^{(n)}$ (see Table I) |
| Step 4 | Compute $d\Omega_n = (d\Omega_{n-1}) \cup \left(\frac{\partial(\Omega_n)}{\partial z}\right)$ <br> IF n=$k_{max}$ go to Step 6 <br> ELSE go to Step2 |
| Algorithm to compute Lie symmetries | |
| Step 5 | Calculate the null space $B = \{^1\xi_z \ ^2\xi_z \ ... \ ^r\xi_z\} = \text{null}(d^{(n)}\Omega_n)$. |
| Step 6 | Analytically solve the following differential equations: <br> $\frac{\partial(^i\phi_z)}{\partial \epsilon_i} = {}^i\xi_z({}^i\phi_z)$, ${}^i\phi_z|_{\epsilon_i=0} = z$, $i = 1,2,...,r$. |
| Algorithm to detect observability | |
| Step 7 | Calculate $r_n = \text{rank}(d\Omega_n)$ |
| Step 8 | IF $r_n = N_z$, observable, ELSEIF $r_n < N_z$, unobservable |

In situations where an input-output mapping is found to be unobservable, it becomes critical to examine the partial observability for each state individually. To address this issue, Maes et al. [37] defined the partial observability for each state: let $d^{(n)}\Omega_n^m$ denote the matrix that is obtained after removing the m-th column from the matrix $d^{(n)}\Omega_n$. If and only if $\text{rank}(d^{(n)}\Omega_n^m) = \text{rank}(d^{(n)}\Omega_n) - 1$, the m-th state variable is n-row observable.

Furthermore, as observed from Eq. (17), the partial observability of a single state implies that this state will not appear in the general expression of Lie symmetries. This observation provides an alternative perspective to handle an unobservable input-output mapping. For instance, the next section will demonstrate that if a state (either dynamic states or unknown parameter) is identified as unobservable, adding a sensor related to the dynamic state or assuming the unknown parameter as known can restore the observability of the state.

**Table III. Algorithm for Detecting Partial Observability**

| | |
| --- | --- |
| Step 1 | Starting from m = 1 |
| Step 2 | Compose $d^{(n)}\Omega_n^m$ by removing the m-th column from $d^{(n)}\Omega_n$, where $m \in 1,2,...,N_z$ |
| Step 3 | Calculate $r_n = \text{rank}(d^{(n)}\Omega_n^m)$ |
| Step 4 | The m-th state is n-row observable if and only if $\text{rank}(d^{(n)}\Omega_n^m) = \text{rank}(d^{(n)}\Omega_n) - 1$ |
| Step 5 | IF m = $N_z$ END <br> ELSE m = m + 1 and go to Step 2 |

**Table I. Various Definitions for the Lie Derivative of Output Function**

| | Input-output mapping | The Lie derivatives of the output function |
|---|---|---|
| (1) [34,46] | $x = \begin{bmatrix} x_t \\ \theta \end{bmatrix}$, $\dot{\theta} = 0$ <br> $\dot{x} = f(x) = \begin{bmatrix} f_t(x) \\ 0_{N_\theta \times 1} \end{bmatrix} = f_a(x) + \sum_{j=1}^{N_u} g_{uj}(x)u_j$ <br> $y = h(x, u, w) = h_0(x)$ | $\Omega_n = \left[ L_{f_a}(\Omega_{n-1}); L_{g_{u1}}(\Omega_{n-1}); \ldots; L_{g_{uN_u}}(\Omega_{n-1}) \right]$, <br> with $\Omega_0 = h_0(x)$ |
| (2) [37,47] | $x^{(n)} = \begin{bmatrix} x \\ w^{(0)} \\ \vdots \\ w^{(n)} \end{bmatrix}$ <br><br> $\dot{x}^{(n)} = f^{(n)}(x^{(n)}, u, w^{(n+1)}) = \begin{bmatrix} f(x, u, \theta, w) \\ w^{(1)} \\ w^{(2)} \\ \vdots \\ w^{(n+1)} \end{bmatrix} = f_a^{(n)}(x^{(n)}) + \sum_{j=1}^{N_u} g_{uj}^{(n)}(x) u_j +$ <br> $\sum_{j=1}^{N_w} S_j^{(n)} w_j^{(n+1)}$ <br> $y = h(x, u, w) = h_0(x) + \sum_{j=1}^{N_u} h_{uj}(x) u_j + \sum_{j=1}^{N_w} h_{wj}(x) w_j$ | $\Omega_n = \left[ L_{f_{xw}^{(n-1)}}(\Omega_{n-1}); L_{g_{u1}^{(n-1)}}(\Omega_{n-1}); \ldots; L_{g_{uN_u}^{(n-1)}}(\Omega_{n-1}) \right]$ <br> where $f_{xw}^{(n-1)} = f_a^{(n-1)}(x^{(n-1)}) + \sum_{j=1}^{N_w} S_j^{(n-1)} w_j^{(n)}$ and $g_{uj}^{(n)}(x) = \begin{bmatrix} g_{uj}(x) \\ 0_{(n+1)N_w \times 1} \end{bmatrix}$, <br> where $f_a^{(n)}(x^{(n)}) = \begin{bmatrix} f_a(x) + \sum_{j=1}^{N_w} g_{w_j}(x) w_j \\ w^{(1)} \\ \vdots \\ w^{(n)} \\ 0_{N_w \times 1} \end{bmatrix}$ and $S_j^{(n)} = \begin{bmatrix} 0_{(N_x + nN_w + j - 1) \times 1} \\ 1_{1 \times 1} \\ 0_{(N_w - j) \times 1} \end{bmatrix}$ <br> Specifically, the initial values are: $\Omega_0 = \begin{bmatrix} h_{xw}^T & h_{u1}^T & \ldots & h_{uN_u}^T \end{bmatrix}^T$, where $h_{xw}(x^{(0)}) = h_0(x) +$ <br> $\sum_{j=1}^{N_w} h_{wj}(x) w_j$, $w_j^{(0)} = w_j$, $f_a^{(0)}(x^{(0)}) = \begin{bmatrix} f_a(x) + \sum_{j=1}^{N_w} g_{wj}(x) w_j \,; 0_{N_w \times 1} \end{bmatrix}$, $S_j^{(0)} = \begin{bmatrix} 0_{(N_x + j - 1) \times 1} \\ 1_{1 \times 1} \\ 0_{(N_w - j) \times 1} \end{bmatrix}$, and <br> $g_{uj}^{(0)}(x) = \begin{bmatrix} g_{uj}(x) \\ 0_{N_w \times 1} \end{bmatrix}$ |
| (3) [36,45] | $x^{(n)} = \begin{bmatrix} x \\ w^{(0)} \\ \vdots \\ w^{(n)} \end{bmatrix}$ <br><br> $\dot{x}^{(n)} = f^{(n)}(x^{(n)}, u, w^{(n+1)}) = \begin{bmatrix} f(x, u, \theta, w) \\ w^{(1)} \\ w^{(2)} \\ \vdots \\ w^{(n+1)} \end{bmatrix}$ <br> $y = h(x, u, w)$ | $\Omega_n = \frac{\partial L_f^{n-1} h}{\partial x} f + \sum_{i=1}^{n} \frac{\partial L_f^{n-1} h}{\partial w^{(i-1)}} w^{(i)} + \sum_{i=1}^{n} \frac{\partial L_f^{n-1} h}{\partial u^{(i-1)}} u^{(i)}$, with $\Omega_0 = h(x, u, w)$ |

\* The definition of $L_S(\Omega)$, indicating the Lie derivative of a function vector $\Omega$ along a vector field $S$, can be found in Appendix II

\* $x^{(0)} \neq x$, $f_a^{(0)} \neq f_a$, $w_j^{(0)} = w_j$

*5.2 Schemes to Destroy Lie Symmetries and Restore Observability*

If an input-output mapping is identified as unobservable, Shi's study [41] has demonstrated that the following three schemes are effective in restoring observability and eliminating Lie symmetries. Additionally, establishing a new mathematical model, such as a simplified one with fewer degrees-of-freedom based on justifiable assumptions, appears to be another option from the authors' understanding.

Scheme 1: Adding new measurement. If there is a new measurement defined as $y_{new} = h_{new}(q, u)$, in order to destroy the ith group of Lie symmetries $^i\phi$, it should satisfy:

$$\frac{dh_{new}(q, u)}{dq} {}^i\xi(q) \neq 0, \qquad i = 1, 2, \dots, r \tag{20}$$

Since the infinitesimals $^i\xi(q)$ have been computed in Eq. (18), we can check if the ith group of Lie symmetries is destroyed by substituting the new measurement function into Eq. (20).

For instance, the most straightforward approach is to measure an unmeasured dynamic state such as displacement or velocity. In Eq. (20), the Jacobian of a newly measured displacement or velocity with respect to the state vector, which contains all displacements and velocities of the system, is clearly a non-zero vector. This vector post-multiplied by an existing infinitesimal $^i\xi(q)$ (also non-zero) results in a non-zero value, satisfying Eq. (20).

Scheme 2: Treating an unknown parameter as known or properly imposing constraints. Since the unknown parameter is included in the augmented state vector, the example above has proven that when a state variable is known, one Lie group of symmetries will be destroyed.

Scheme 3: Transforming the system's state vector into a new state vector, which should not exceed the original one's dimension. In this scenario, the transformation and the ith group of symmetries can be expressed as:

$$q_T = T(q), \qquad {}^i\phi_T = T({}^i\phi) \tag{21}$$

To destroy the ith group of Lie symmetries, the new vector should be independent of $\epsilon_i$, yielding to:

$$\frac{dT(q)}{dq} {}^i\xi(q) = 0, \qquad i = 1, 2, \dots, r \tag{22}$$

where we can check if the ith group of Lie symmetries is destroyed by substituting the transformation function into Eq. (22).

*5.3 Connection with the Condition of a Unique Solution of Linear Equation Group*

This connection has been clarified in many previous studies. For the sake of completeness, a linear system, which satisfies the Lie derivative definition in Table I (1), is taken as an example.

$$\begin{aligned}\dot{x} &= Ax + Bu \\ y &= Cx\end{aligned} \tag{23}$$

where A, B, C are constant matrices, and $f_a(x) = Ax$, $g_u(x) = B$ and $h_0(x) = Cx$.

Applying the algorithm yields to following computation steps:
　Initial: Set $k_{max} = N_x-1$, n=0, z = x, $\Omega_0 = Cx$, $d\Omega_0 = C$.
　Iteration 1: $\Omega_1 = [CAx; CB]$ and $d\Omega_1 = [C; CA]$.
　Iteration 2: $\Omega_2 = [CA^2x; 0_{1\times N_x}; CAB; 0_{1\times N_x}]$, simplified as $\Omega_2 = [CA^2x; CAB]$, such that $d\Omega_2 = [C; CA; CA^2]$.
　⋮
　Iteration $N_x$-1: $\Omega_{N_x-1} = [CA^{N_x-1}x; CA^{N_x-2}B]$ and $d\Omega_{N_x-1} = [C; CA; CA^2; \dots; CA^{N_x-1}]$.

It is seen that the Jacobian matrix is exactly the observability matrix of the linear system, namely $d\Omega_{N_x-1} = \mathcal{O}(A, C)$, defined in many textbooks as:

$$\begin{bmatrix} y(t_0) \\ \dot{y}(t_0) \\ \ddot{y}(t_0) \\ \vdots \\ y^{(N_x-1)}(t_0) \end{bmatrix} = \begin{bmatrix} C \\ CA \\ CA^2 \\ \vdots \\ CA^{N_x-1} \end{bmatrix} x(t_0) = \mathcal{O}(A,C)x(t_0) \qquad (24)$$

which indicates that the initial condition $x(t_0)$ can be uniquely determined if and only if the linear observability matrix has full rank, i.e., $\text{rank}\mathcal{O}(A,C)=N_x$.

Furthermore, since a derivative with respect to time in a continuous system corresponds to a time shift in the corresponding discrete-time system, we can transform the above equation into:

$$\begin{bmatrix} y(0) \\ y(1) \\ y(2) \\ \vdots \\ y(N_x-1) \end{bmatrix} = \begin{bmatrix} C_d \\ C_d A_d \\ C_d A_d^2 \\ \vdots \\ C_d A_d^{N_x-1} \end{bmatrix} x(0) = \mathcal{O}(A_d,C_d)x(0) \qquad (25)$$

where $C_d = C$ and $A_d = e^{A\Delta t}$, according to the discretization of the continuous system. Similarly, the initial condition $x(0)$ can be uniquely determined if and only if the linear observability matrix has full rank, i.e., $\text{rank}\mathcal{O}(A_d,C_d)=N_x$.

Therefore, from a linear algebra viewpoint, the Jacobian matrix in nonlinear observability analysis corresponds to the coefficient matrix of Eq. (25), whose rank determines the uniqueness of solutions. The existence of Lie symmetries corresponds to infinitely many solutions of Eq. (25) when $\text{rank}\mathcal{O}(A_d,C_d) < N_x$.

## 6. Case study

### 6.1 Confirmation of the Consistency with Previous Studies

An example used in previous studies [36,37,41] for illustrative purposes can demonstrate the consistency between different Lie derivatives of the output function and the association between unobservable states and the action of Lie symmetries on these states. This example approximates an isolated model with two degrees of freedom. The isolation layer's mechanical behavior is idealized as a nonlinear spring element with a stiffness hardening effect, while the superstructure remains in its elastic range. Viscous damping is considered. A measured force u and an unmeasured force w are applied to the base and first stories, respectively, with two accelerometers installed at the two stories.

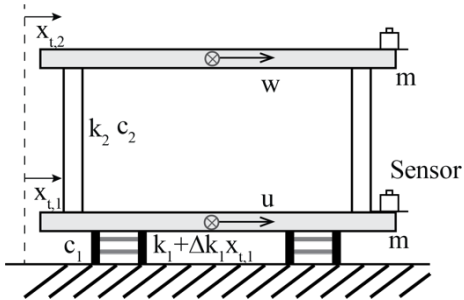

**Figure 3.** Illustrative Example

The input-output mapping of the system, see Eq. (1), can be established as:

$$\dot{x}_t = \frac{d}{dt}\begin{bmatrix} x_{t,1} \\ x_{t,2} \\ \dot{x}_{t,1} \\ \dot{x}_{t,2} \end{bmatrix} = \begin{bmatrix} \dot{x}_{t,1} \\ \dot{x}_{t,2} \\ \dfrac{-(k_1+\Delta k_1 x_{t,1})x_{t,1} + k_2(x_{t,2}-x_{t,1}) - c_1\dot{x}_{t,1} + c_2(\dot{x}_{t,2}-\dot{x}_{t,1}) + u}{m} \\ \dfrac{k_2(x_{t,1}-x_{t,2}) + w}{m} \end{bmatrix} \qquad (26)$$

$$y = \begin{bmatrix} \ddot{x}_{t,1} \\ \ddot{x}_{t,2} \end{bmatrix} = \begin{bmatrix} \dfrac{-(k_1 + \Delta k_1 x_{t,1})x_{t,1} + k_2(x_{t,2} - x_{t,1}) - c_1\dot{x}_{t,1} + c_2(\dot{x}_{t,2} - \dot{x}_{t,1}) + u}{m} \\ \dfrac{k_2(x_{t,1} - x_{t,2}) + c_2(\dot{x}_{t,2} - \dot{x}_{t,1}) + w}{m} \end{bmatrix} \quad (27)$$

Given the unmeasured input, the Lie derivatives of the output function in Table I (2) (3) can be used. Table I (3) has been used in a previous study [41] to compute the general expression of Lie symmetries. To validate consistency, the definition of Table I (2) is used.

In this case, the system identification problem involves estimating the dynamic states, four unknown parameter and the unmeasured force, as: $[x, w] = [x_{t,1} \; x_{t,2} \; \dot{x}_{t,1} \; \dot{x}_{t,2} \; k_1 \; \Delta k_1 \; k_2 \; m \; w]$. The rest are assumed to be known. The input-output mapping is extended to the following form:

$$\dot{x}^{(n)} = f^{(n)}(x^{(n)}) + g_u^{(n)}(x)u + S_w^{(n)} w^{(n+1)}, \qquad y = h_0(x) + h_u(x)u + h_w(x)w \quad (28)$$

where $x^{(n)} = [x \; w^{(0)} \; \dots \; w^{(n)}]$, and

$$f^{(n)}(x^{(n)}) = \left\{ \begin{bmatrix} \dot{x}_{t,1} \\ \dot{x}_{t,1} \\ \dfrac{-(k_1 + \Delta k_1 x_{t,1})x_{t,1} + k_2(x_{t,2} - x_{t,1}) - c_1\dot{x}_{t,1} + c_2(\dot{x}_{t,2} - \dot{x}_{t,1})}{m} \\ \dfrac{k_2(x_{t,1} - x_{t,2}) + c_2(\dot{x}_{t,2} - \dot{x}_{t,1})}{m} \\ 0_{4\times 1} \\ w^{(1)} \\ \vdots \\ w^{(n)} \\ 0 \end{bmatrix} + \begin{bmatrix} 0_{3\times 1} \\ \dfrac{1}{m} \\ 0_{4\times 1} \end{bmatrix} w \right\}$$

$$g_u^{(n)}(x) = \begin{bmatrix} 0_{2\times 1} \\ 1/m \\ 0_{(5+n+1)\times 1} \end{bmatrix}, \qquad S_w^{(n)} = \begin{bmatrix} 0_{(8+n)\times 1} \\ 1 \end{bmatrix}$$

and

$$h_0(x) = \begin{bmatrix} \dfrac{-(k_1 + \Delta k_1 x_{t,1})x_{t,1} + k_2(x_{t,2} - x_{t,1}) - c_1\dot{x}_{t,1} + c_2(\dot{x}_{t,2} - \dot{x}_{t,1})}{m} \\ \dfrac{k_2(x_{t,1} - x_{t,2}) + c_2(\dot{x}_{t,2} - \dot{x}_{t,1})}{m} \end{bmatrix}, \; h_u(x) = \begin{bmatrix} \dfrac{1}{m} \\ 0 \end{bmatrix}, \; h_w(x) = \begin{bmatrix} 0 \\ \dfrac{1}{m} \end{bmatrix}$$

First, the general expression of Lie symmetries is computed. Performing the algorithm up to order n=6 shows that a null space basis exists in the Jacobian matrix $d\Omega_6$, as:

$$^1\xi_{q^{(6)}} = \left[ \underbrace{\underbrace{1, \dfrac{k_1 + k_2}{k_2}, 0, 0, -2\Delta k_1, 0, 0, 0}_{^1\xi_x}, \underbrace{k_1, 0, 0, 0, 0, 0}_{^1\xi_w}}_{^1\xi_{x^{(5)}}}, \underbrace{0}_{^1\xi_{w^{(6)}}} \right]^T \quad (29)$$

Substituting it into Eq. (8) leads to the following expression:

$$\frac{d({}^1\phi_{x^{(6)}})}{d\epsilon_1} = \frac{d}{d\epsilon_1}\begin{bmatrix}{}^1\phi_{x,1}\\ \vdots\\ {}^1\phi_{x,8}\\ {}^1\phi_w\\ {}^1\phi_{w^{(1)}}\\ \vdots\\ {}^1\phi_{w^{(6)}}\end{bmatrix} = \begin{bmatrix}1\\ \frac{{}^1\phi_{x,5}+{}^1\phi_{x,7}}{{}^1\phi_{x,7}}\\ 0\\ 0\\ -2\,{}^1\phi_{x,6}\\ 0\\ 0\\ 0\\ {}^1\phi_{x,5}\\ 0\\ 0\\ 0\\ 0\\ 0\\ 0\end{bmatrix}, \quad \text{where } {}^i\phi_{x^{(6)}}|_{\epsilon_1=0} = \begin{bmatrix}x\\ w\\ w^{(1)}\\ \vdots\\ w^{(6)}\end{bmatrix} \quad (30)$$

Solving the first order differential equation leads to the expression as:

$${}^1\phi_x = \begin{bmatrix}x_{t,1}+\epsilon_1\\ x_{t,2}+\dfrac{k_1+k_2}{k_2}\epsilon_1 - \dfrac{\Delta k_1}{k_2}\epsilon_1^2\\ \dot{x}_{t,1}\\ \dot{x}_{t,2}\\ k_1 - 2\Delta k_1 \epsilon_1\\ \Delta k_1\\ k_2\\ m\end{bmatrix}, \quad {}^1\phi_w = w + k_1\epsilon_1 - \Delta k_1 \epsilon_1^2, \quad {}^1\phi_{w^{(k)}} = w^{(k)}\ (i=1,\dots,6) \quad (31)$$

The computed expression of Lie symmetries aligns with that in Shi's study [41] using the definition of Table I (3). Conversely, the observability analysis result in this case is presented in Fig. 4a. The left plot depicts the target rank and the calculated rank of the Jacobian matrix at each step, while the right plot illustrates the partial observability of each state at each step. It is observed from the left plot that after step 3, the difference between the target rank and the calculated rank is 1, implying the existence of 1 group of Lie symmetries. From the right plot, it is observed that the unobservable states are $\{x_{t,1}, x_{t,2}, k_1, w\}$, which correspond to the state variables acted upon by Lie symmetries, as seen in $\epsilon_1$ in Eq. (31).

Three examples are shown for restoring observability. In the first, the unobservable parameter $k_1$ is assumed to be known. The new 'output function' can be treated as $h_{new} = k_1$, and subsequently substituted into Eq. (20) resulting in Eq. (32), a non-zero value indicating the destruction of Lie symmetries. Similarly, adding a displacement transducer at the base floor to measure the unobservable state $x_{t,1}$ will lead to the same effect. The observability can also be restored in the second and third cases where the accelerometer at the base layer is replaced by a displacement transducer at the superstructure or a displacement transducer at the base layer, respectively, as shown in Figs. 4c and 4d.

$$\begin{bmatrix}0 & \dots & \overset{k_1}{\overset{\frown}{1}} & \dots & 0\end{bmatrix}{}^1\xi_{q^{(6)}} = k_1 \quad (32)$$

We have shown a unified understanding for the association between Lie symmetries and observability analysis. The difference between the target rank and the calculated rank is the number of existing Lie groups of symmetries, and the action of these Lie symmetries lies on the unobservable states. In the remainder of this paper, only the observability analysis will be used to detect the uniqueness of solutions of a specified input-output mapping.

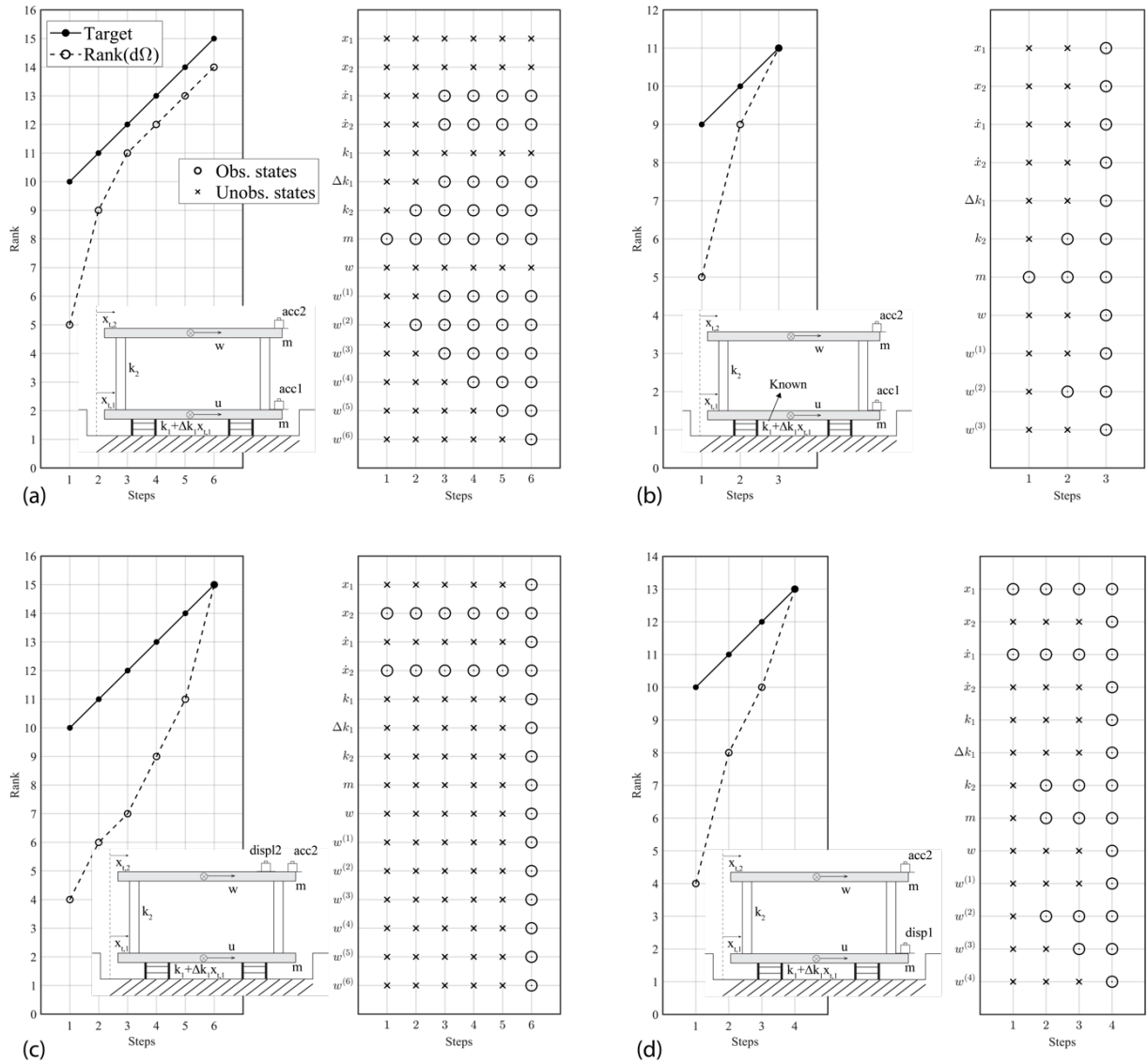

**Figure 4.** Observability Analysis Results: (a) Two Accelerometers (b) Known Unknown Parameter (c) Superstructure Accelerometer & Displacement Transducer (d) Superstructure Accelerometer & Base Layer Displacement Transducer.

*6.2 Structures with Different Damping Systems and Observability Analysis Results*

This section uses three examples to illustrate the unified algorithm's use and highlight the distinct demands for sensor configurations in a 5-story building with different damping systems under various circumstances.

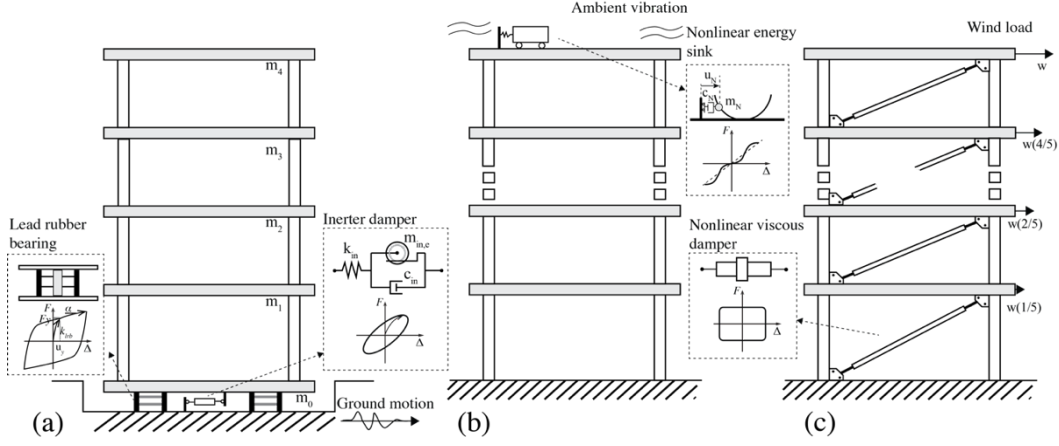

**Figure 5.** Three Building Configurations: (a) Seismic Isolation and Inerter Dampers (b) Nonlinear Energy Sink, Ambient Vibration (c) Nonlinear Dampers, Unmeasured Wind Loads.

In the first case, the building is equipped with isolation techniques using lead rubber bearings combined with inerter dampers under earthquake action, as shown in Fig. 5a. The inerter damper, as proposed in Ikago's study [14], comprises a linear spring in series with a viscous damper and an inerter in parallel. This damper amplifies resistant force by harnessing the rotational motion of a relatively small mass. The two ends of the inerter mass and its rotation angle satisfy the relationship, $x_{in}(t) = r\theta(t)$, where r is the radius of gyration, leading to the inerter force as:

$$I_{in}\ddot{\theta} = f_{in}r, \qquad f_{in} = \frac{I_{in}}{r}\ddot{\theta} = \frac{1}{2}m_{in}\frac{R^2}{r^2}\ddot{x}_{in} = m_{in,e}\ddot{x}_{in} \qquad (33)$$

where $I_{in}$ is the moment of inertia of the inerter, and $m_{in,e}$ is the equivalent mass of the inerter in terms of its translational motion. Consequently, the governing motion of the entire inerter damper can be expressed as:

$$f_{in} = m_{in,e}\ddot{x}_{in} + c_{in}\dot{x}_{in} = k_{in}(x_0 - x_{in}) \qquad (34)$$

The lead rubber bearing (LRB) [48,49], a popular seismic isolation technique for buildings, bridges, and other infrastructures, is represented using the differentiable Bouc-Wen model [50] because of its continuous and smooth properties that facilitate successive differentiation operations in the observability analysis algorithm.

$$f_{srdi} = \alpha k_{lrb}x_0 + (1-\alpha)k_{lrb}u_y z, \qquad \dot{z} = \frac{1}{u_y}\dot{x}_0\big(1 - (\tanh(\rho z)z)^{n_{lrb}}(\gamma + \beta\tanh(\rho z \dot{x}_0))\big) \qquad (35)$$

where $k_{lrb}$ denotes the pre-yielding stiffness, $\alpha$ denotes the ratio of post-yielding stiffness to pre-yielding stiffness, $u_y$ is the yielding displacement, and z is an evolutionary hysteresis displacement. In addition, $\gamma$, $\beta$, $n_{lrb}$ are coefficients of the common Bouc-Wen model, and $\rho$ is a sufficiently large value.

Given that the main structure is strategically designed to remain in its elastic range, the governing equation of the system can be expressed by a shear building model as:

$$M\ddot{x} + C\dot{x} + Kx = -M1_{N_x \times 1}\ddot{u}_g + Tf_{in} \qquad (36)$$

where T is a location vector. Combining Eqs. (33) - (36) can solve the dynamic response of the structure.

Assuming all dynamic states and parameters as unknown, namely $\{m_0, \ldots, m_4\}$ $\{c_1, \ldots, c_4\}$ $\{k_1, \ldots, k_4\}$ $\{k_{lrb}, \alpha, u_y\}$ $\{m_{in}, k_{in}, c_{in}\}$, the acceleration of input ground motions is measured by an accelerometer installed on the ground. Considering two sensor layouts—an accelerometer or a displacement transducer mounted on the second floor— observability analysis results based on the defined Lie derivative in Table I (1) show that both configurations yield an observable input-output mapping, suggesting that all dynamic states, including unknown parameters, can be uniquely inferred in theory.

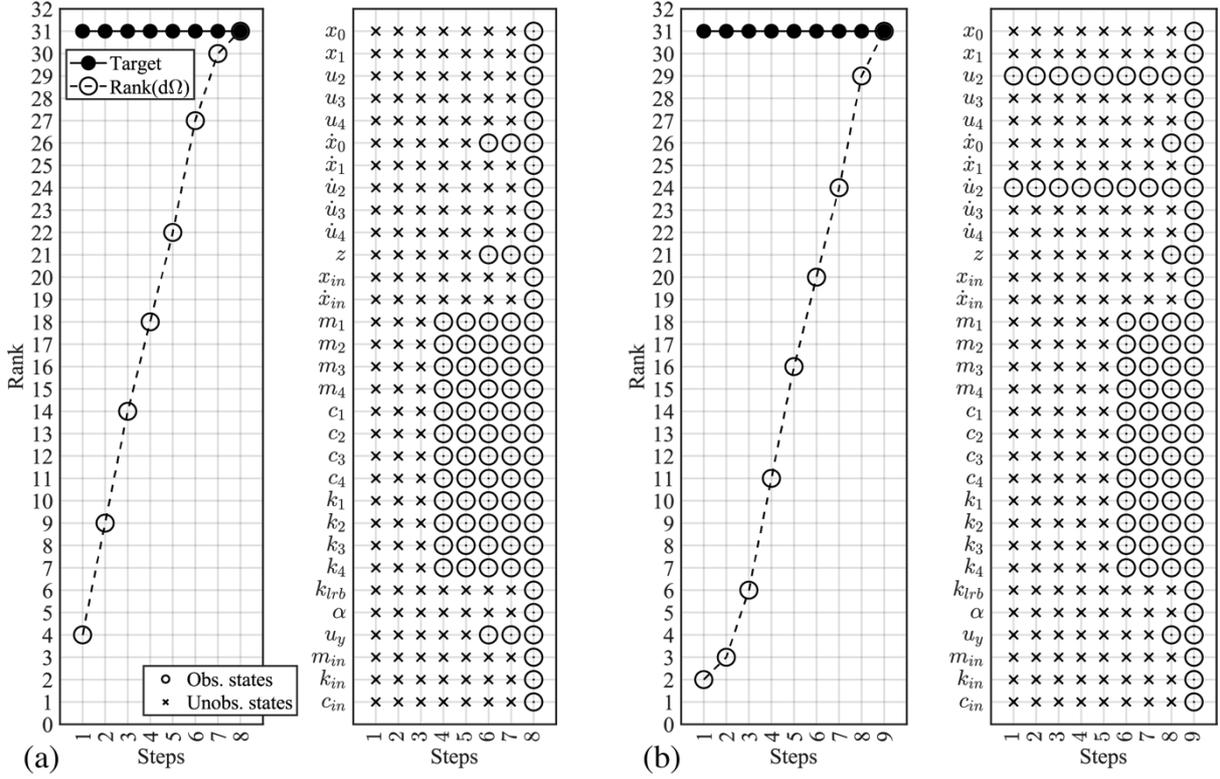

**Figure 6.** Observability Analysis for Isolated Building with Inerter Dampers: (a) Accelerometer on Second Floor, (b) Displacement Transducer on Second Floor.

The second case involves a track nonlinear energy sink (NES) [51], analogous to tuned mass dampers, mounted on the building's top under ambient vibration without inputs, as shown in Fig. 5b. The movement of an auxiliary mass along a track generates nonlinear restoring forces determined by the shaped track, $h(x_N) = a_N x_N^4$, as per

$$f_N(x_N, \dot{x}_N, \ddot{x}_N) = \{[h'(x_N)]^2 \ddot{x}_N + h'(x_N)h''(x_N)\dot{x}_N + h'(x_N)g\}m_N = (16a_N^2 x_N^6 \ddot{x}_N + 48a_N^2 x_N^5 \dot{x}_N^2 + 4a_N x_N^3 g)m_N \tag{37}$$

where $x_N$ is the horizontal displacement of NES relative to its track, and $m_N$ denotes the auxiliary mass. The governing equation of the building featuring a NES on it top can be characterized as:

$$\begin{aligned} M\ddot{x} + C\dot{x} + Kx &= -M1_{N_x \times 1}\ddot{u}_g + T(f_N + c_N \dot{x}_N) \\ m_N \ddot{x}_N + c_N \dot{x}_N + f_N(x_N, \dot{x}_N, \ddot{x}_N) &= -m_N(\ddot{x}_5 + \ddot{u}_g) \end{aligned} \tag{38}$$

where $c_N$ represent all type of damping in the track NES, and $\ddot{x}_5$ is the acceleration of the 5[th] floor relative to the ground.

In system identification, all dynamic states including $\{x_N, \dot{x}_N\}$ and parameters of the track NES $\{m_N, c_N, a_N\}$ are considered. Considering two sensor layouts—accelerometers installed throughout all floors and two displacement transducers measuring displacements of the 4[th] and 5[th] floors—observability analysis results based on the defined Lie derivative in Table I (1) are presented in Fig. 7, demonstrating that either acceleration measurement or displacement measurement can result in an observable input-output mapping.

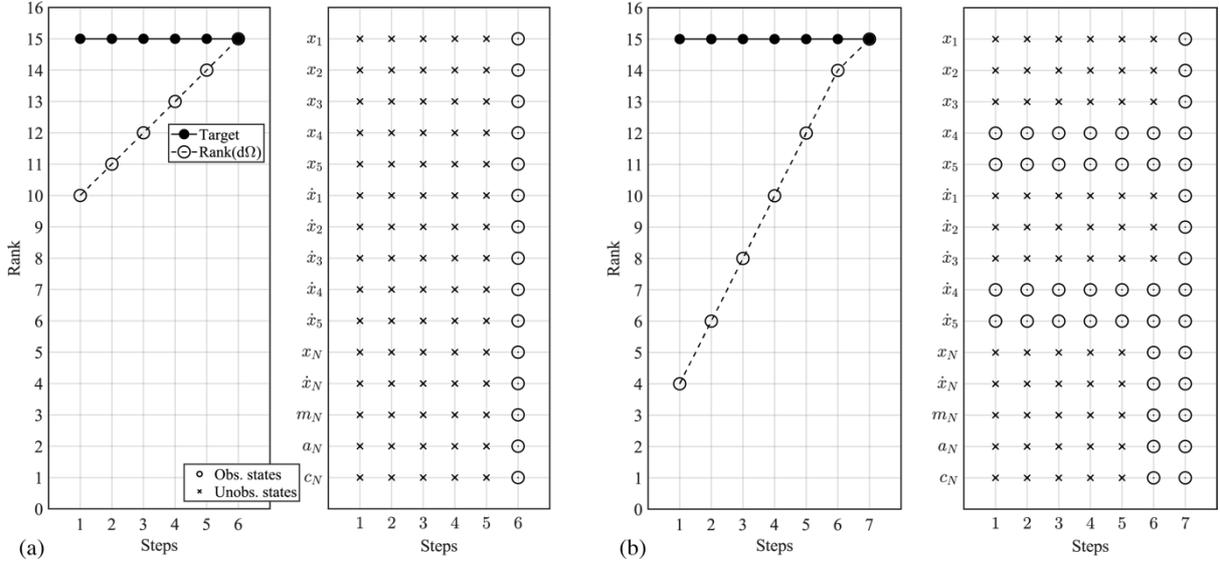

**Figure 7.** Observability Analysis for Building with Top-Floor NES, Ambient Vibration: (a) Accelerometers on All Floors, (b) Displacement Transducers on 4th and 5th Floors.

The third example features nonlinear viscous dampers [52] placed throughout the building under unmeasured wind loads with linearly increasing magnitude, as seen in Fig. 5c. The dampers' nonlinear restoring forces are approximated by the fractional power law of velocity model as per

$$f_{V,i} = C_i |\dot{x}_{V,i}|^\alpha \text{sign}(\dot{x}_{V,i}) \approx C_i |\dot{x}_{V_i}|^\alpha \tanh(\rho \dot{x}_{V_i}), \qquad i = 1,2,\dots,5 \qquad (39)$$

where $\dot{x}_V$ represents the relative velocity between the two ends of the damper, and $C_i$ is damping coefficient. The tanh function, similar to the differentiable Bouc-Wen model Eq. (35), is used to approximate the signum function. Similar to Eq. (36), incorporating the nonlinear restoring forces of the damper into the governing equation can solve dynamic responses of the structure. In addition, due to the discontinuity at the zero point of the absolute function, the absolute operator $|\cdot|$ has been omitted in the observability analysis. The inclusion of the absolute operator results in two branches yielding the same observability analysis outcome.

For system identification, all dynamic states, dampers' parameters $\{C_1, \dots, C_5\}$, and the unmeasured wind load $\{w\}$ are considered. Two sensor layouts are examined, either accelerometers or displacement transducers installed on all floors. The observability analysis results, based on the defined Lie derivative in Table I (2), are shown in Fig. 8. The mapping based on all floors' displacement measurement is unobservable, with 3 Lie groups of symmetries acting on states $C_1$, w and its higher order derivatives. However, all floors' acceleration measurement suggests the existence of one Lie symmetry, and the partial observability analysis result shows that all quantities of interest are observable, except the higher order of w. From a practical standpoint, this mapping can be considered observable.

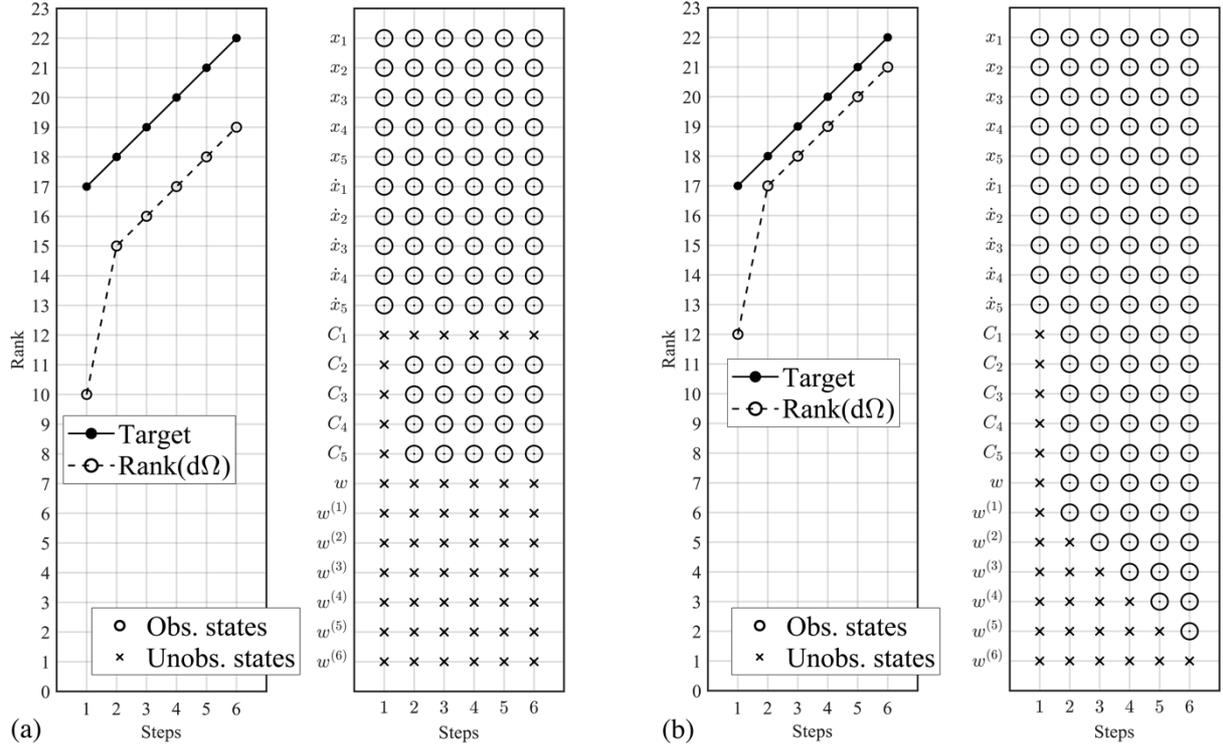

**Figure 8.** Observability Analysis for Building with Nonlinear Dampers on All Floors: (a) Displacement Measurements (b) Acceleration Measurements.

*6.3 Joint State-Parameter Estimation for an Observable Input-Output Mapping Via Unscented Kalman Filter*

The first case is employed as an example to execute system identification. Among diverse system identification methodologies, the Unscented Kalman Filter (UKF) has proven effective [53] and has been used in various problems, such as time-varying parameters and noise [54], as well as unknown inputs [55,56]. The UKF, based on unscented transform, generates sigma points to balance approximation accuracy and computational cost when estimating the system's state probability distribution.

The design philosophy for seismic isolation structures aims to focus the main structure's deformation and energy dissipation on the isolation layer, which minimizes the seismic response's reliance on other parameters. The dynamic states and unknown parameters considered in this section include:

- $\{x_0 \cdots x_4 \; \dot{x}_0 \cdots \dot{x}_4 \; z \; x_{in} \; \dot{x}_{in} \; k_{lrb} \; \alpha \; u_y \; m_{in} \; k_{in} \; c_{in}\}$

The building model's structural information is tabulated in Table IV, and the isolation layer parameters are detailed in Table V. The inerter damper's parameters are aligned with Pan's optimal design formula [57].

**Table IV.** Benchmark Structure's Information

| Floor | Height | Mass (ton) | Stiffness (kN/mm) | Damping ratio |
|---|---|---|---|---|
| 4 | 3.80 | 1800 | 2488 | 0.03 |
| 3 | 4.40 | 1807 | 1939 | 0.03 |
| 2 | 4.40 | 1928 | 2038 | 0.03 |
| 1 | 5.45 | 2335 | 1760 | 0.03 |
| 0 (Isolation layer) | 1.00 | 3057 | - | - |

**Table V.** Design Parameters of Lead Rubber Bearings and Inerter Dampers

| LRB | $k_{lrb}$ (kN/mm) | $f_y$ (kN) | $u_y$ (mm) | $\alpha$ | $n_{lrb}$ | $\beta$ | $\gamma$ | $\rho$ |
|---|---|---|---|---|---|---|---|---|
| | 135 | 5400 | 40 | 0.2 | 2 | 0.75 | 0.8 | 100 |
| Inerter damper | $m_{in} = \tau/\sum m$ | | | $k_{in} = \kappa k_{lrb}$ | | | $c_{in} = 2\zeta\sqrt{k_{lrb}\sum m}$ | |
| | $\tau = 0.1$ | | | $\kappa = 0.12$ | | | $\zeta = 0.013$ | |

Three cases with estimated parameters randomly deviating from true values by -50% to 200% are considered. For each case, the remaining parameters randomly vary from their true values within ±20% uncertainties. The 1994 Kobe earthquake's ground acceleration at the Takatori station is used as measured inputs. An accelerometer on the second-floor slab measures the seismic response as output $y = \ddot{x}_{2,abs}$. The unknown parameters are updated by their normalized coefficients $\alpha_\eta(t)$, reflecting changes relative to the true value:

$$\eta_t = \alpha_\eta(t)\eta_0 \tag{40}$$

where $\eta_0$ represents a parameter's true value. For the UKF filter setting, the system noises covariance aligns with a $1 \times 10^{-4}\text{rms}(\ddot{x}_{2,abs})$ variance for all dynamic states and a $2 \times 10^{-5}$ variance for all estimated parameters' normalized coefficients. Measurement noises covariance matches a $2 \times 10^{-2}\text{rms}(\ddot{x}_{2,abs})$ variance. The initial error covariance for estimated parameters is $2 \times 10^{-4}$. The measurement sampling frequency is 100 Hz, but the measurement used for performing the UKF has linearly interpolated 30 steps between two measured points due to the highly nonlinear characteristics of the system. More information can be found in previous studies by the authors [58,59] and the Appendix III.

**Table VI.** Initial Values of Parameters

| Random variations within ±20% on: | | $m_0, ..., m_4, c_1, ..., c_4, k_1, ..., k_4, n_{lrb}, \beta, \gamma, \rho$ | | | | |
|---|---|---|---|---|---|---|
| $\alpha_\eta(0)$ | $k_{lrb}$ | $\alpha$ | $u_y$ | $m_{in}$ | $k_{in}$ | $c_{in}$ |
| Case I | 0.58 | 0.79 | 1.18 | 0.60 | 0.77 | 1.24 |
| Case II | 1.88 | 0.64 | 0.85 | 0.61 | 1.78 | 0.77 |
| Case III | 1.96 | 1.85 | 0.75 | 1.678 | 1.98 | 0.78 |

The simulation results, created from design values, serve as reference or true values. Figure 9 showcases the estimation results for absolute accelerations and relative displacements throughout the building. The acceleration responses and base layer displacement response align well with their true values. Superstructure interstory drifts also match the reference response, except for minor discrepancies in Case I. Although clear drifts occur early due to the imposed ±20% uncertainties, the superstructure's maximum drift is less than 0.01 cm, a negligible amount.

Figures 10 and 11 display the estimation results for the isolation bearing parameters, damping devices, and corresponding force-displacement relationships. The estimated parameters with varying initial errors ultimately converge to their true values with acceptable precision. Additionally, the restored force-displacement relationships of individual devices, from the lead rubber bearing and the complete inerter damper to its subcomponents - the viscous damper and the inerter mass, align well with their true relationships. This comparison provides a direct method for managers and manufacturers to confirm the performance of individual devices under operational conditions, offering a valuable perspective for maintenance.

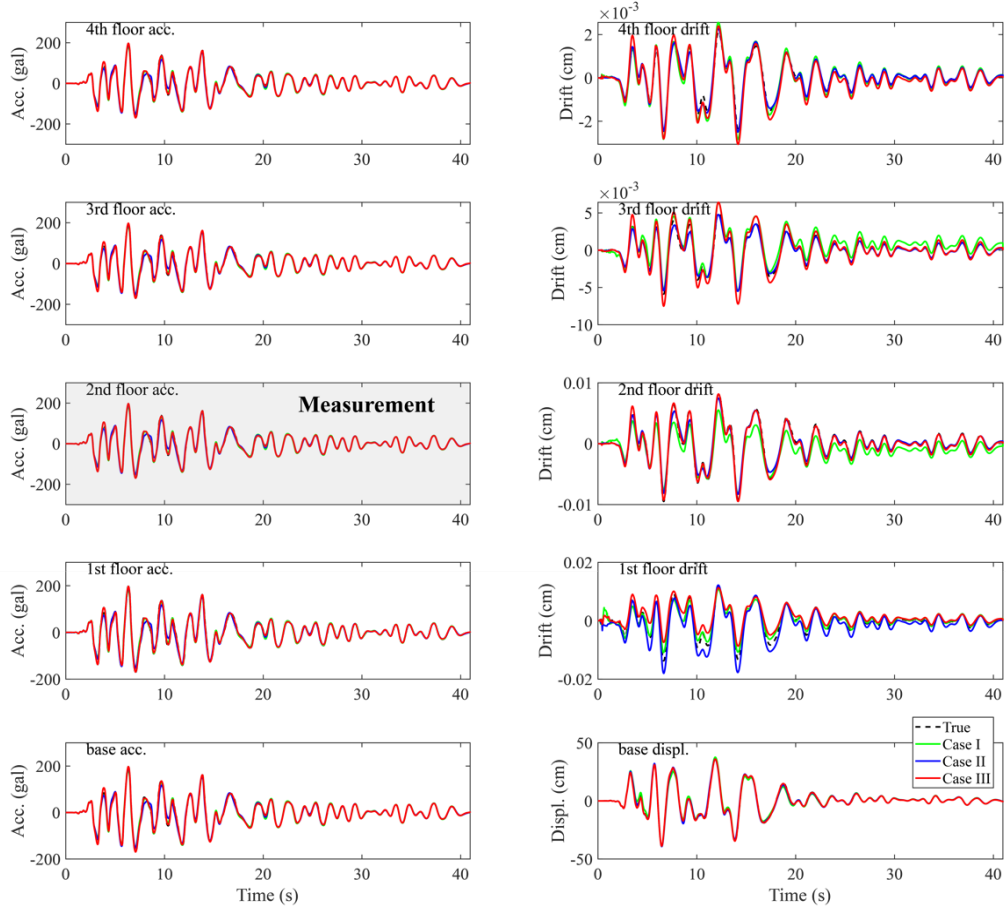

**Figure 9.** Estimation Results for Structural Responses, Absolute Accelerations and Inter-Story Displacements

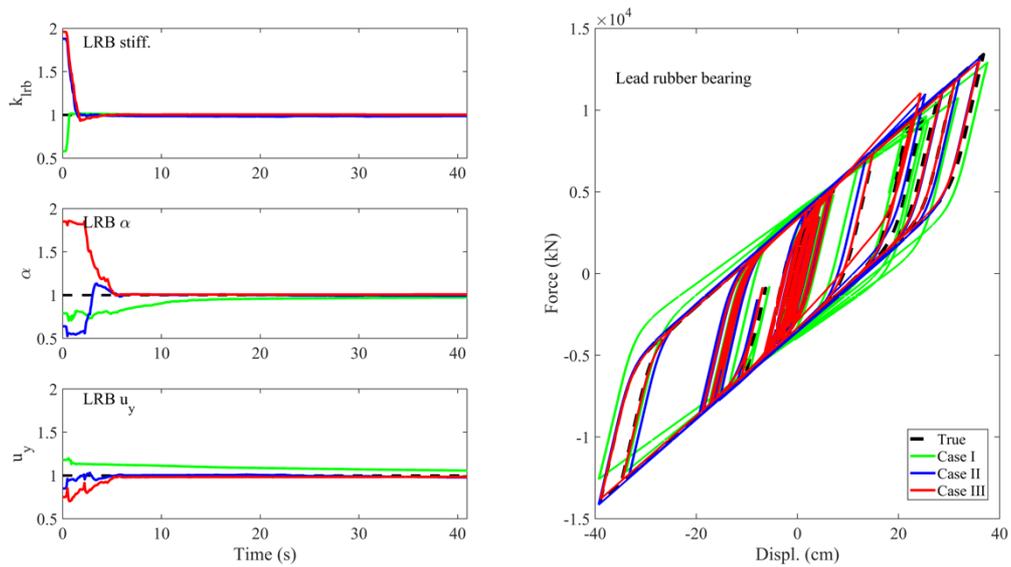

**Figure 10.** Estimation Results for the Lead Rubber Bearing

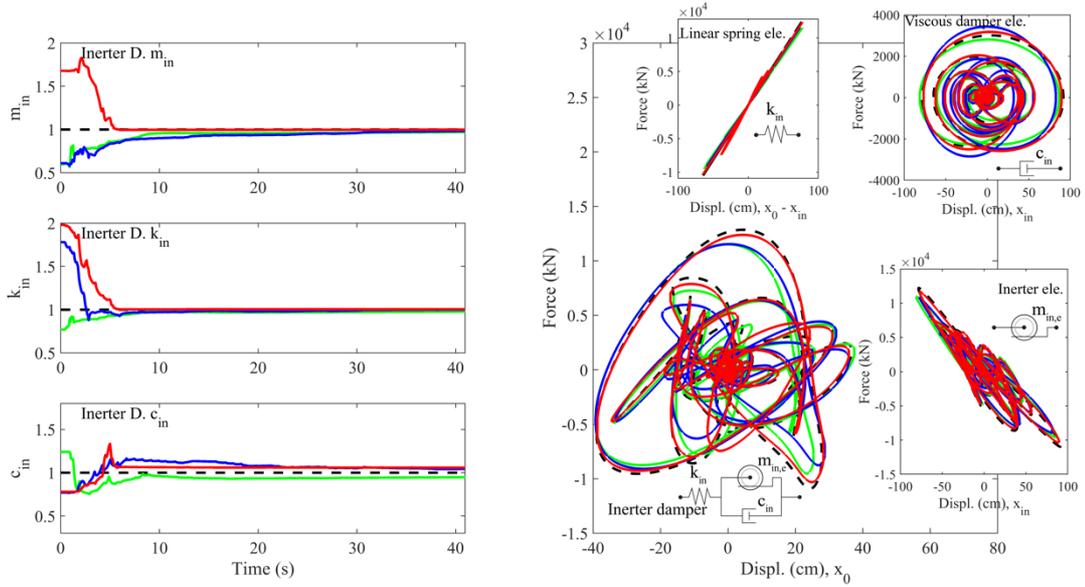

**Figure 11.** Estimation Results for the Inerter Damper

## 7. Conclusion

The increasing integration of supplementary damping systems in modern civil engineering structures necessitates performance assessment under operational conditions, thereby invoking the need for system identification techniques. A key challenge in this context involves constructing an input-output mapping that assures unique solutions for system identification methods. Addressing this issue, this study offers a unified understanding to computing Lie symmetries and observability analysis from an engineering standpoint. Specifically, a flexible unified algorithm is presented, the adaptability of which hinges on substituting the Lie derivatives' definition of the output function to meet the diverse range of practical applications. The main contributions can be summarized as follows:

1) A unified understanding for the association between Lie symmetries and observability analysis is presented. The difference between the target rank and the calculated rank is the number of existing Lie groups of symmetries, and the action of these Lie symmetries lies on the unobservable states.
2) The methodology is exemplified with three cases of a 5-story building equipped with different damping systems subjected to distinct situations, including seismic action, ambient vibration, and unmeasured wind loads. The observability analysis results underline some unique demands for sensor configurations in practical system identification.
3) The effectiveness of observable input-output mapping in parameter convergence and error minimization is demonstrated through the application of the unscented Kalman filter for joint state-parameter estimation of a seismic isolation building model with lead rubber bearings and inerter dampers, even though the number of sensors has been minimized.

**Appendix I: Some Definitions and Properties of Lie Symmetries.**

We present a concise overview of Lie symmetries. For a more comprehensive understanding, please refer to Oliveri's review study [42].

Definition A.1 (Groups of transformations): Consider a domain $D \subseteq \mathbb{R}^N$ and a subset $S \subseteq \mathbb{R}$. The set of transformations

$$x^* = \phi(x, \epsilon), \qquad \phi: D \times S \to D \qquad (A.1)$$

depending the parameter $\epsilon$, form a one-parameter group of transformations on $D$, if the following conditions hold:
1) For each value of the parameter $\epsilon \in S$, the transformations are one-to-one onto $D$;
2) $S$ with the law of composition $\mu$ forms a group with identity $e$;
3) $\phi(x, e) = x$, for all $x \in D$;
4) $\phi(\phi(x, a), b) = \phi(x, \mu(a, b))$, for all $x \in D$, and for all $a, b \in S$.

Definition A.2 (Lie group of transformations): The group of transformation (A.1) defines a one-parameter Lie group of transformations, if in addition to satisfying the axioms of the previous definition:
1) $\epsilon$ is a continuous parameter, i.e., S is an interval in $\mathbb{R}$;
2) $\phi$ is infinitely differentiable $C^\infty$ with respect to x in D and an analytic function of x in S;
3) $\mu(a, b)$ is an analytic function of a and b, for all $a, b \in S$.

Due to the demand for the analyticity of the group operation $\mu$, it is always feasible to reparametrize the Lie group such that the group operation aligns with the ordinary sum in $\mathbb{R}$. By expanding (A.1) in powers of $\epsilon$ around $\epsilon = 0$, we obtain (within a neighborhood of $\epsilon = 0$):

$$x^* = x + \epsilon \frac{\partial \phi(x, \epsilon)}{\partial \epsilon}\Big|_{\epsilon=0} + \epsilon^2 \frac{\partial^2 \phi(x, \epsilon)}{\partial \epsilon^2}\Big|_{\epsilon=0} + \cdots = x + \epsilon \frac{\partial \phi(x, \epsilon)}{\partial \epsilon}\Big|_{\epsilon=0} + O(\epsilon^2) \quad (A.2)$$

Be defining

$$\xi(x) = \frac{\partial \phi(x, \epsilon)}{\partial \epsilon}\Big|_{\epsilon=0} \quad (A.3)$$

the transformation

$$x^* = x + \epsilon \xi(x) \quad (A.4)$$

outlines the infinitesimal transformation of the Lie group of transformations, where the components of $\xi(x)$ are the so-called infinitesimals of Eq. (A.1).

First Fundamental Theorem of Lie: The Lie group of transformations (A.1) corresponds to the solution of the initial value problem for the first-order differential equations:

$$\frac{dx^*}{d\epsilon} = \xi(x^*), \quad x^*(0) = x \quad (A.5)$$

The First Fundamental Theorem of Lie indicates that the infinitesimals encompass the required information for characterizing a one-parameter Lie group of symmetries.

**Appendix II: Lie Derivative of a Function Vector along a Vector Field**

Without losing generality, the Lie derivative of a function vector $\Omega(x) = [\omega_1(x); \ldots; \omega_{N_\Omega}(x)]$ along a vector field $\mathcal{S} = [s_1; \ldots; s_{N_S}]$ is formulated as:

$$L_\mathcal{S}(\Omega(x)) = \frac{\partial \Omega(x)}{\partial x} \cdot \mathcal{S} \quad (A.6)$$

where $\cdot$ denotes the post-multiplication of a matrix with a column vector, and the operation $d\Omega(x) = \frac{\partial \Omega}{\partial x}$ for a column vector $\Omega(x)$ with respect to a column vector $x = [x_1; \ldots; x_{N_x}]$ refers to the Jacobian:

$$\frac{\partial \Omega}{\partial x} = \begin{bmatrix} \frac{\partial \omega_1}{\partial x_1} & \cdots & \frac{\partial \omega_1}{\partial x_{N_x}} \\ \vdots & \ddots & \vdots \\ \frac{\partial \omega_{N_\Omega}}{\partial x_1} & \cdots & \frac{\partial \omega_{N_\Omega}}{\partial x_{N_x}} \end{bmatrix} \quad (A.7)$$

The m-th order Lie derivative of a function vector $\Omega(x)$ along a series of fields $\mathcal{S}_j, j = 1, 2, \ldots, m$, is defined as:

$$L_{\mathcal{S}_m} \circ \ldots \circ L_{\mathcal{S}_2} \circ L_{\mathcal{S}_1}(\Omega(x)) \quad (A.8)$$

where $\circ$ denotes the successive application of the Lie derivative. The zero-order Lie derivative is equivalent to $\Omega(x)$ itself.

**Appendix III: Generalized Formulation of an Adaptive Unscented Kalman Filter**

As Bayesian-type methods, the Kalman filter treats the augmented system states as random variables and produces the corresponding posterior probabilistic estimates. Analogous to Eq. (4), the dynamics of the augmented system are then governed by $\dot{q} = F(q, u, v_F)$ and $y = H(q, u, v_H)$, where $v_F \sim \mathcal{N}(0, Q)$ and $v_H \sim \mathcal{N}(0, R)$ denote zero-mean white Gaussian processes with the covariance matrix Q and R representing the system and measurement noise, respectively. As a nonlinear variant of the Kalman filter, the unscented Kalman filter (UKF) [31,60] propagates the first two moments of the states through suitably selected sigma points and the corresponding weights. The most general formulation of the UKF concatenates the system and measurement noise with the state vector q to form the further augmented vector $q^a$ with dimensions of $N = N_x + N_{v_F} + N_{v_H}$, entering the system and measurement equations in a nonlinear manner, $q^a = [q; v_F; v_H]$, where the semicolon denotes the concatenation of vectors. In addition, the adaptive scheme to update the system and measurement noise covariances is shown to improve the trackability and robustness of the Kalman filter.

Assuming time $t = i\Delta t$, the estimate of the augmented state vector is known as $\hat{q}_i^{a,+} = [\hat{q}_i^+; 0; 0]$, where the symbol '+' denotes the posterior estimate (in contrast, '−' denotes the prior prediction), and the covariance matrix is known as $P_{q_i^{a,+}} = \text{diag}\left(P_{q_i^+}, Q_i, R_i\right)$. The algorithm of the general UKF with an adaptive scheme is summarized in Table A1. Further details are presented in our previous studies, e.g., Refs. [50,58,59,61].

**Table A1.** Algorithm of the generalized adaptive UKF (AUKF)

| |
| --- |
| Prediction update step |
| a. Generate $2N + 1$ sigma points as: $q_{i,0}^{a,+} = \hat{q}_i^{a,+}$, $q_{i,j}^{a,+} = \hat{q}_i^{a,+} + \left(\sqrt{(N+\kappa)P_{q_i^{a,+}}}\right)_i$, $q_{i,N+j}^{a,+} = \hat{q}_i^{a,+} - \left(\sqrt{(N+\kappa)P_{q_i^{a,+}}}\right)_i$, $j = 1,2,\ldots,N$ where $\sqrt{\cdot}$ is the matrix square root and $(\cdot)_i$ denotes the i-th column of the matrix, $\kappa$ is a scaling parameter and can be any number as long as $N + \kappa > 0$. [62] |
| b. Propagate each sigma point $q_{i,j}^{a,+} = [q_{i,j}^{q,+}; q_{i,j}^{w,+}; q_{i,j}^{v,+}]$ through the nonlinear system equation to update the time as: $q_{i+1,j}^{q,-} = q_{i,j}^{q,+} + \int_{i\Delta}^{(i+1)\Delta t} F(q_{i,j}^{q,+}, u_i, q_{i,j}^{w,+}) dt$ |
| c. Calculate the prior estimate and the corresponding covariance matrix as: $\hat{q}_{i+1}^- = \sum_{j=0}^{2N} W_j q_{i+1,j}^{q,-}$, $P_{q_{i+1}^-} = \sum_{j=0}^{2N} W_j (q_{i+1,j}^{q,-} - \hat{q}_{i+1}^-)(q_{i+1,i}^{q,-} - \hat{q}_{i+1}^-)^T$, where the weighting coefficients for the sigma points are calculated as $W_0 = \frac{\kappa}{N+\kappa}$, and $W_j = \frac{1}{2(N+\kappa)}$, $j = 1,2,\ldots,2N$ |
| Measurement update step |
| a. Generate $2N + 1$ sigma points again using the updated covariance matrix as: $P_{q_{i+1}^{a,-}}$ |
| b. Evaluate the measurement of each sigma point as: $\mathcal{Y}_{i+1,j} = F(q_{i+1,j}^{q,-}, u_k, q_{i+1,j}^{v,-})$ |
| c. Calculate the estimated measurement, the innovation covariance matrix, and the cross covariance as: $\hat{y}_{i+1} = \sum_{j=0}^{2N} W_j \mathcal{Y}_{i+1,j}$, $P_{y_{i+1}} = \sum_{j=0}^{2N} W_j (\mathcal{Y}_{i+1,j} - \hat{y}_{i+1})(\mathcal{Y}_{i+1,j} - \hat{y}_{i+1})^T$, and $P_{qy_{i+1}^-} = \sum_{j=0}^{2N} W_j (q_{i+1,j}^{q,-} - \hat{x}_{i+1}^-)(\mathcal{Y}_{i+1,j} - \hat{y}_{i+1})^T$ |
| d. Calculate the Kalman gain as: $\mathbb{K}_i = P_{qy_{i+1}^-}(P_{y_{i+1}})^{-1}$ |
| e. Calculate the posterior estimate and the covariance matrix as: $\hat{q}_{i+1}^+ = \hat{q}_{i+1}^- + \mathbb{K}_i(y_{i+1} - \hat{y}_{i+1})$, $P_{q_{i+1}^+} = P_{q_{i+1}^-} - \mathbb{K}_i P_{y_{i+1}} \mathbb{K}_i^T$ |
| Adaptive adjustment step |
| a. Update the noise covariances based on a stochastic approximation scheme [63] as: $Q_{i+1} = (1 - \alpha_Q)Q_i + \alpha_Q \mathbb{K}_i(y_i - \hat{y}_i)(y_i - \hat{y}_i)^T \mathbb{K}_i^T$, and $R_{i+1} = (1 - \alpha_R)R_i + \alpha_R(y_i - \hat{y}_i)(y_i - \hat{y}_i)^T$, where $\alpha_R = 1/30$, $\alpha_Q = 1/30$ are the adaptive coefficients |
| Repeat the above procedure for the next step $i + 1$. |